

\def\svec#1{\skew{-2}\vec#1}
\magnification=1200
\hoffset=-.1in
\voffset=-.2in

\vsize=7.5in
\hsize=5.6in
\tolerance 10000

\baselineskip 12pt plus 1pt minus 1pt
\pageno=0
\centerline{{\bf VERY SMALL STRANGELETS}\footnote{*}{This
 work is supported in part by funds
provided by the U. S. Department of Energy (D.O.E.) under contract
\#DE-AC02-76ER03069.}}
\vskip 24pt
\centerline{E.~P.~Gilson and R.~L.~Jaffe}
\vskip 12pt
\centerline{\it Center for Theoretical Physics}
\centerline{\it Laboratory for Nuclear Science}
\centerline{\it and Department of Physics}
\centerline{\it Massachusetts Institute of Technology}
\centerline{\it Cambridge, Massachusetts\ \ 02139\ \ \ U.S.A.}
\vskip 1.5in
\centerline{Submitted to: {\it Physical Review Letters\/}}
\vfill
\centerline{ Typeset in $\TeX$ by Roger L. Gilson}
\vskip -12pt
\noindent CTP\#2173\hfill February 1993
\eject
\baselineskip 24pt plus 2pt minus 2pt
\centerline{\bf ABSTRACT}
\medskip
We study the stability of small strangelets by employing a simple model of
strange matter as a gas of non-interacting fermions confined in a bag.  We
solve the Dirac equation and populate the energy levels of the bag one quark
at a time.  Our results show that for system parameters such that strange
matter is unbound in bulk, there may still
exist strangelets with $A<100$ that are
stable and/or metastable.  The lifetime of these strangelets may be too small
to
detect in current accelerator experiments, however.
\vfill
\eject
\noindent{\bf I.\quad INTRODUCTION}
\medskip
\nobreak
With the advent of heavy ion colliders, it will soon be possible to search for
stable or metastable lumps of quark matter with\footnote{}{*For the sake of
simplicity, throughout this Letter we assign the strange quark a strangeness
of $+1$.} $S^*\sim A\sim
10-30$.$^1$  The possible stability of strange quark
matter (``strange matter'') in bulk was pointed out by Witten in
1984,$^2$ and since then, there have been numerous attempts to predict
the properties of strange matter in bulk and in finite lumps
(``strangelets'').$^3$  These studies generally only apply to baryon
numbers much larger than those accessible in heavy ion colliders.  Our
intention in this letter is to present some qualitative information on very
small strangelets obtained
from a very elementary model.  Our model includes only
quark kinetic energy, the Pauli principle and confinement.  It cannot tell
us anything about important issues like the overall energy scale, or
equivalently the bulk stability, of strange matter.  It does, however,
illustrate potentially interesting effects such as shell closures, ``loading''
 and unloading of strangeness and isolated ``islands of stability'' in the
$(S,A)$ plane.  None of the details of our predictions should be taken very
seriously; they would undoubtedly be changed in more sophisticated models.
Nevertheless, the types of phenomena which occur in our model may well
persist in others.

No one knows how to model quark matter in QCD accurately.  Lattice
simulations are as yet unable to cope with systems at non-zero chemical
potential.  Models of bulk strange matter have confined quarks in a bag and
included residual gluon interactions perturbatively.$^{4,\,5}$  Surface
effects were included for large strangelets by including surface
modifications of the quark density of states as well as Coulomb
effects.$^6$  The resulting Thomas-Fermi like model is only valid for
strangelets with radii very large compared to the natural length scale of
the system, $B^{-1/4} \sim 1-2$~fm.  For typical strange matter
densities, a strangelet with baryon number $A \sim 200$  has a radius of
only $5-6$~fm, so only for such a large baryon number does
the model of Ref.~[6] become reliable.
We model a small strangelet as a gas of
non-interacting fermions confined
in a bag.  Rather than approximate the density of states, we instead fill
the bag energy levels sequentially, obeying the exclusion principle,
minimizing the energy (for each $A$) and adjusting the bag radius so the
quark pressure balances the vacuum pressure, $B$.  The free parameters we
use are the energy per baryon in bulk, $\epsilon_{b}$, the mass of the
strange quark $m_{s}$, and of course, the baryon number, $A$.  We ignore
residual perturbative QCD interactions following Ref.~[6] where it was argued
(on the basis of Ref.~[5]) that the effect of the interactions can be largely
absorbed into a redefinition of the overall energy scale parameterized by
$\epsilon_{b}$.  We also ignore Coulomb corrections.  This should be a
good approximation because $Z$ is very small for small $A$, $Z<\!\!<A$.
Both of these effects should be included in future calculations.

Both the quantum numbers and energetics of small strangelets show
characteristic regularities reminiscent of atomic physics.  We see that
``shell'' effects are extremely important when filling a bag with quarks:
the rate of change of the energy per baryon with A changes dramatically
near shell closures and leads to enhanced stability.  We find that
there exist small regions of $A$ in which strangelets are stable even for
system parameters such that strange matter is not bound in bulk.  We also
observe that the strangeness of the most stable strangelet is an erratic
function of $A$.  Strange and non-strange quark energy levels cross as a
function of the bag radius $R$.  When, as happens, a non-strange level
dives below a strange level, the strange level ``unloads'' into the
non-strange one, dropping strangeness by as much as $\Delta S = -6$ from one
 value of $A$ to the next.  This phenomenon is similar to the filling and
emptying of inner $d$-orbitals in the periodic table.  Finally, we find that
the spatial distribution of strangeness is not uniform throughout a strangelet.
Because they are less relativistic, strange quarks are concentrated in the
interior and depleted on the surface.  This phenomenon is related to the
quark mass dependence of the surface tension in strange matter.

In our model, we describe the system as a Fermi gas under constant pressure.
  First, we fix the strange quark mass and the bag radius, $R$.  After
determining the energy eigenvalues for the quarks, we fill the bag and vary
its radius until the quark pressure at the surface equals the
vacuum pressure, $B$.  This is equivalent to minimizing the total energy at
constant pressure, $B$.  Values of $B$ were taken from fits to bulk strange
 matter as described in Ref.~[6].

There have been previous attempts to use similar models to study very small
strangelets. In Ref.~[7], the strangelets were constructed by filling the bag
energy levels with the strangeness ratio held fixed.  In the published work,
$S/A$ was fixed at $0.7$ which is far from the optimal value for most $A$.
 Fixing the strangeness ratio artificially prevents the system from finding a
minimum energy configuration.  Thus, the model of Ref.~[7] is not adequate for
studies of stability.

More recently, Madsen$^8$ attempted to study very small strangelets using
the asymptotic expansion of the density of states including the curvature
term in order to achieve higher accuracy.
In fact, we find that there is no region in $A$ in which the curvature term
gives a useful correction to the density of states.  At large $A$, the
curvature term is
negligible.  At small $A$, where it would be expected to be important, other
even lower-order corrections are even more important.  This is clearly
displayed in Fig.~1 where we plot the integral of the density of states,
$N(k)$, for a light quark as a function of $k = \sqrt{E^2-m^2}$ in a rigid
spherical cavity.  For comparison we plot the first two terms in an asymptotic
expansion for $N(k)$ for large systems, obtained by integrating the density of
states, $\rho(k)$,
$$\rho(k) = {g\over2\pi^2}\left[ k^{2}V -{\pi kS\over 4}
\left(
1- {2\over \pi}\tan^{-1}{k\over m}\right)+\ldots \right]\ \ ,\eqno(1)$$
where $V$ and $S$ are the volume and surface area of the bag, respectively.
The first omitted term in Eq.~(1) is proportional to the surface integral of
the average curvature ($\oint d^2s{1\over 2} \left( 1/R_1 + 1/R_2\right)$,
where $R_1$ and $R_2$ are the principal curvatures at each point) and
suppressed by ${\cal O}(1/k^2)$ relative to the volume term.
At low $k$, $N(k)$ is a noisy function of $k$ reflecting
the details of the eigenvalue spectrum of the Dirac operator in a cavity.  At
large $k$, $N(k)$ is well-approximated by the integral of Eq.~(1).
There does not
seem to be a significant intermediate region in which a ${\cal O}(1/k^2)$
(curvature) correction is significant.  We conclude that the asymptotic
expansion of the density of states is not useful to study very small
strangelets.
\goodbreak
\bigskip
\noindent{\bf II.\quad STRANGELETS AS NON-INTERACTING DIRAC FERMIONS}
\medskip
\nobreak
Since we are dealing with light quarks confined to a small bag, we must of
course consider them as relativistic particles.  We write down the Dirac
equation and the appropriate boundary condition,
$$\eqalignno{ ( \svec{\alpha} \cdot \svec{p} + \beta m ) \Psi &= E \Psi\ \
,\qquad \hbox{for}\ r<R\ \ ,&(2)\cr
i\hat{r}\cdot\svec{\gamma}\Psi &= \Psi\ \ ,\qquad \hbox{for}\ r=R
\ \ .&(3)\cr}$$
This boundary condition ensures that there is no probability flux leaving
the bag ($\svec{\jmath} \cdot \hat{r} = 0$).  Note that the wavefunction and
the
density, $\Psi^{\dagger} \Psi$, need not go to zero on the boundary, whereas
for the non-relativistic case, the wave function must vanish at the boundary.
  This implies that the more massive, hence less relativistic, strange quarks
 will tend to shy away from the boundary of the bag.

Once the Dirac equation is solved with this  boundary condition and
geometry, we obtain expressions for the eigenfunctions
and transcendental equations for the eigenvalues.  We take the energy,
momentum, and mass to be $\omega$, $k$, and $m$ respectively.  We define
$\alpha = \omega R$, $x=kR$, and $\lambda = mR$, so that
$\alpha^{2}=x^{2}+\lambda^{2}$.  Thus, the eigenvalue equation reads
$$\sqrt{\alpha + \lambda}\,f_{\kappa} = -\sqrt{\alpha - \lambda}\,f_{\kappa -1}
\eqno(4)$$
 where $f$ is the spherical Bessel function regular at the origin, and
$\kappa=- \ell-1$ for $j=\ell+1/2$ and $\kappa=\ell$ for $j=\ell-1/2$.  Using
this definition of $\kappa$, we can re-express the angular momentum/color
degeneracy, $3(2j+1)$ as $6|\kappa|$.
  The normalized wave function is
$$\Psi = \sqrt{ {x^{2}\over  R^{3}(2\alpha(\alpha + \kappa)+
\lambda)f^2_\kappa(x) }}
\left(
\matrix{{\displaystyle{if_\kappa \left( {xr\over R}\right) \phi^\ell_{jm}}}
\cr\noalign{\vskip 0.2cm}
{\displaystyle{{x\over \alpha+\lambda} f_{\kappa-1} \left( {xr\over R}\right)
{\svec\sigma} \cdot \hat r \phi^\ell_{jm}}} \cr}\right) \ \ .\eqno(5)$$
The eigenvalues are a function of the product $mR$, the mass
of the particle times the radius of the bag.  For the up and down quarks, the
 mass is taken to be zero.

As described in Ref.~[6], we determine the bag constant and an estimate of the
radius of the bag
as a function of $\epsilon_{b}$, $m_{s}$, and $A$ by studying the bulk limit,
$A\to\infty$.  In bulk
equilibrium, the Fermi seas for the three quark species must have the same
Fermi
energy or chemical potential:
$\mu = \mu_{\rm up} = \mu_{\rm down} = \mu_{\rm strange}$.  $\mu$
is the change in total
energy due to the addition of a quark.  We obtain the number of
particles/volume $n_{a}$ for $a = u$, $d$, and $s$
by integrating the $k^2V$ term in  the density of states, Eq.~(1).
For the massless $u$ and $d$ quarks, $n_{u,d} = (\mu^{3}/\pi^{2})$ for the
strange quark, $n_{s} = (\mu^{3}\cos^{3}\theta/\pi^{2})$, where
$\sin \theta = (m/\mu)$.  In bulk, the baryon number is
$A=(1/3)\sum_a n_{a}V$. The total energy of the bag is
$E=\sum_a \mu_{a}n_{a}V$.  By using these two expressions, we find that
$\epsilon_{b} = 3\mu$.  As noted in Ref.~[6], the surface tension is
positive.  Thus, to minimize energy, the shape of the strangelet will be
spherical.  By inverting the relation between $A$ and $n$, we obtain a first
estimate of the radius of the bag as a function of
$\epsilon_{b}$, $m_{s}$ and $A$.
$$R = \left( {9\pi A\over 4}\left( 2 + \left[ {\sqrt{\epsilon^2_{b} -
9m^2}\over \epsilon_{b}}\,\right]^3\right)^{-1}\right)^{1/3} {3\over
\epsilon_{b}}\  \ .\eqno(6)$$
This is the radius of a lump of strange matter in which all surface effects
are ignored and is used as a first approximation to the
actual radius that will balance the quark pressure against the vacuum
pressure.  The equilibrium condition on the volume gives us the equation for
$B$, $B= - \sum \Omega_{a}\equiv$ quark pressure.  Using the equations for
$\Omega_{a}$ from Ref.~[6], we get
$$
B = {2\left( {\displaystyle{\epsilon_{b}\over 3}}\right)^4 \over
4\pi^2}+ {1\over 4\pi^2} \left[ {\epsilon_{b}\over 3} \sqrt{
\left( {\epsilon_{b}\over 3}\right)^2 - m^2} \left( \left(
{\epsilon_{b} \over 3}\right)^2 - {5\over 2} m^2\right)
+{3\over 2} m^4 \ln
{ {\displaystyle{\epsilon_{b}\over 3}} + \sqrt{
\left( {\displaystyle{\epsilon_{b}\over 3}}\right)^2 - m^2 } \over
m}\,\right]\ \ .\eqno(7)$$
Once the quark mass has been chosen and a first approximation to the radius
has been determined, we calculate the energy levels by
solving the transcendental equation numerically.  We then adjust the radius,
 and recalculate the energy levels, until the total energy is minimized.
Once a strangelet thus is created, we can read off its energy per baryon,
strangeness, and radius directly.

We have performed a variety of checks on this calculation.  First, we
have calculated the number of states with ``momentum'' less than $k$
$(k= \sqrt{\omega^{2}-m^{2}})$.  This function, $N(k)$, should be
approximated by the integral of the asymptotic expansion of Eq.~(1) for large
$k$.  This check is shown in Fig.~1 where it is clear that our model
reproduces the asymptotic result and the surface correction.  Second, we have
checked that
the energy per baryon and strangeness per baryon also converge to the
bulk values as $A\rightarrow \infty$.  These checks reassure us that our
calculation has been performed correctly.

We now turn to issues of stability and composition of strangelets.  If a
strangelet is not in flavor equilibrium, it can decay via weak semileptonic
decays, weak radiative decays, and electron capture all of which do not change
baryon number.  Other modes of decay such as fission, alpha decay, weak and
strong neutron decays, and  strong $\Lambda$, $\Sigma$, $\Xi$, $\Omega$ decays
reduce baryon number by one or more units.  Our strangelets are already in
equilibrium at a given $A$ and thus are only subject to alpha and strong or
weak
 neutron decays, strange baryon decays, and fission.  We check the
stability of our strangelets against alpha decay, fission, and the other
baryon decays noted.

A strong neutron decay will occur if the difference in energy between two
strangelets of the same strangeness but $\Delta A=-1$ is greater then $m_{n}$.
  Similarly, a weak neutron decay is possible if the energy difference
between two strangelets of $\Delta S=-1$ and $\Delta A=-1$ is greater than
$m_{n}$.  For the $\Lambda$, $\Sigma$, $\Xi$, $\Omega$ decays,
we have $\Delta A=-1$
 and $\Delta S=-1,-1,-2,-3$,  respectively.
We calculated these energies and discovered where stable
regions exist within our model.
\goodbreak
\bigskip
\noindent{\bf III.\quad RESULTS}
\medskip
\nobreak
For small $A$, the dynamics are as follows.  Given the choice between massive
 and massless particles, we opt to fill the bag with less energetic massless
particles first.  We continue to add massless particles until we build up a
large enough Fermi sea so that it becomes energetically favorable to add a
strange quark to the system.  Soon, it again becomes favorable to add
non-strange quarks to the system.  One might expect that strange and
non-strange levels will fill in an alternating sequence.  However, Fig.~2a
shows that this is not the case.
This is because the massive quark energy levels change at a
different rate with respect to the radius
than the massless quark energy levels do.
Energy levels can cross, and strange levels that have been filled may
suddenly empty out into nonstrange levels.  This can be seen on the
 data for $\epsilon_{b} = 950$~MeV,  $m_{s} = 150$~MeV, where a level
crossing occurs at $A = 30$ and the strangelets become stable until the next
nonstrange level begins to fill at $A = 36$ (see Fig.~2a).

Our results indicate that within our model, there exist stable and metastable
strangelets
for various system parameter values.  We generated strangelets of baryon
numbers 1 -- 100 for various values of $m_{s}$ and $\epsilon_{b}$.
We find that for $\epsilon_{b} < 930$~MeV, which is the value of $\epsilon$ for
$^{56}_{26}$Fe, there exist many stable
strangelets.  The stability of strangelets
with several choices of $\epsilon_b$ and $m_s$ is displayed in Fig.~3.  Those
species noted in the figure are stable against single baryon emission.
Whenever the
 slope of the $\epsilon (A)$ curve is negative enough, the decrease in energy
due to emission of a particle is not enough to offset the increase in energy
due to the slope.  Thus, for small $A$, it is frequently energetically
unfavorable for a strangelet to decay via emission of a baryon.  Many of the
smaller strangelets, however, are subject to fissioning into several $\Lambda$
hyperons and a nucleus, or simply dissolving into $\Lambda$ hyperons and
neutrons.  This mode is a strong decay, but its rate will be
suppressed by several orders of magnitude due to the unlikeliness of the
quarks simultaneously arranging themselves into the decay products.  The
suppression is difficult to estimate, however, because we are dealing with a
collective, many-particle effect.  The astute reader will observe that some
quasistable species occur in regions where the slope of $\epsilon(A)$ is
positive (see, for example, Fig.~3d near $A=60$).  In this region the
strangeness charge between most stable species with $A$ and $-1$ is $\Delta
s=-3$ (see Fig.~2a) requiring $\Omega^-$ emission which is energetically
forbidden.  Neutron emission requires $\partial\epsilon/\partial A$ {\it at
fixed strangeness\/} to be positive.  In the region of concern, direct
calculation shows $\partial E/\partial A\big|_s$ to be negative.

Another interesting effect that can be seen is the phenomenon of shell
closures.  The first level to fill is a $1s_{1/2}$ level where $\kappa=-1$.
This level may hold
six quarks of each flavor.
At every occurrence of a shell closure, the $\epsilon (A)$ curve
takes a noticeable dip.  This generates a large slope for $\epsilon (A)$ and
thus, stable regions in the neighborhood of a closed shell.  This is similar
to atomic physics where shell closures produce more stable, less chemically
reactive elements.  Shell closures can be seen at $A = 4$, 6, 14, 18, 22,
$\ldots$ (see Fig.~3).  These particular values correspond to a non-strange
$s_{1/2}$-shell, a strange $s_{1/2}$-shell, a non-strange $p_{3/2}$-shell, a
non-strange $p_{1/2}$-shell, and a strange $p_{3/2}$-shell, respectively.
The locations of these shell closures are a function of the self-consistent
``potential,'' in which the quarks are bound --- in this case, the bag.
Therefore, the precise values should not be taken too seriously.

The most surprising results are uncovered when we examine values of
$\epsilon_{b} > 930$~MeV.  Specifically, looking at
$\epsilon_{b} = 950$, 970~MeV, $m_{s} = 150$~MeV,
we see that there still
exist islands of stability against single baryon decay
(see Fig.~3).  This is interesting because the
failure of terrestrial searches to find stable strange matter suggests that
strange matter in bulk may well be unstable.$^9$  Our
results indicate that even though this may be the case, there is still a
chance of detecting small strangelets in the laboratory provided the strong
decay into light nuclei and several hyperons or complete dissolution
does not proceed too rapidly to
allow the produced strangelets to reach the detector before decaying.
These islands of
stability persist until $\epsilon_{b} \sim 1000$~MeV, $m_{s}=150$~MeV.

The charge systematics of light strangelets are important for experimenters.
In bulk, we expect roughly equal numbers of $u$-, $d$- and $s$-quarks, thus
$Z/A<\!\!<1$.  Even for nuclei, where $Z\sim A$, Coulomb effects are not
important for small $A$.  For small strangelets we ignore them.  The possible
charges of small strangelets are determined by which shells are filled, and
which one is currently filling.  Figure~2b shows that the allowed charges for
strangelets as a function of baryon number is a complex function that
reflects the nature of the shell filling process.  Throughout our region of
interest, the charge remains relatively small (and occasionally negative) in
comparison to $A$, so we are justified in neglecting the Coulomb energy
contribution.

We also plotted the spatial density for the quarks in the strangelets.
 As is guaranteed by the Dirac equation, the heavier, less relativistic,
strange quarks in fact have a distribution that is concentrated closer to the
 center of the strangelet
than the up and down quarks (see Fig.~4).  This is a reflection of
the boundary condition imposed.  By requiring that no probability flux leave
the bag rather than requiring that $\Psi=0$, a relativistic quark may have
a non-zero density, $\Psi^{\dagger} \Psi$, at the
boundary.  As the mass of the particle increases, we approach the
non-relativistic limit where the boundary condition becomes $\Psi = 0$.
Thus, strange (heavier) quarks are depleted near the surface.

We have shown that the energetics associated with shell closures
are likely to be important in the study of very small strangelets.  Our
admittedly crude method brings out this aspect of the system that is not seen
when the
smoothed density of states is employed.  Our results are consistent with those
obtained for large $A$.  We therefore conclude that metastable
strange matter may be
found in small lumps.  The suppressed strong decay into a nucleus (or many
neutrons) and $\Lambda$ hyperons, might render it difficult to detect, however.
One characteristic that would identify a strangelet is
 its unusual charge/mass ratio.  The charge is typically small since flavor
equilibrium favors charge neutrality even for relatively small $A$.
\vfill
\eject
\centerline{\bf REFERENCES}
\medskip
\item{1.}J..Barrette {\it et al.\/}, {\it Phys. Lett.\/} {\bf 252}, 550
(1990); J. Sandweiss, {\it et al.\/}, proposal to RHIC to be submitted.
\medskip
\item{2.}E.~Witten, {\it Phys.~Rev.\/} {\bf D30}, 272 (1984).
\medskip
\item{3.}For an introduction see, {\it Strange Quark Matter in Physics and
Astrophysics\/}, J.~Madsen and P.~Haensel, eds., {\it Nucl. Phys. B\/}
(Proceedings Supplement), {\bf 24B} (1991) December 1991.
\medskip
\item{4.}S.~Chin and A.~K.~Kerman, {\it Phys.~Rev.~Lett.\/} {\bf 43}, 1242
(1979).
\medskip
\item{5.}E.~Farhi and R.~L.~Jaffe, {\it Phys.~Rev.\/} {\bf D30}, 2379 (1984).
\medskip
\item{6.}M.~Berger and R.~L.~Jaffe, {\it Phys.~Rev.\/} {\bf C35}, 213 (1987).
\medskip
\item{7.}C.~Greiner, D.-H.~Rischke, H.~St\"oker and R.~Koch, {\it
Phys.~Rev.\/} {\bf D38}, 2797 (1988).
\medskip
\item{8.}J.~Madsen, {\AA}rhus University preprint, June 1992.
\medskip
\item{9.}E. G. Blackman and R. L. Jaffe, {\it Nucl. Phys. B.\/} {\bf 324}, 205
(1989).
\vfill
\eject
\item{Fig.~1:}The scaled number of states as a function of $k$ compared with
the asymptotic expansion including
the surface correction. Here, $m_s=150$~MeV, $R=1$, $g=6$.
\medskip
\item{Fig.~2a:}The strangeness as a function of $A$ for the most stable
species, illustrating the
``unloading'' of strange quarks into non-strange energy levels as strange and
non-strange energy levels cross.  Here $\epsilon_b=950$~MeV, $m_s=150$~MeV.
\medskip
\item{Fig.~2b:}The allowed range of charges for strangelets as a function of
$A$.
\medskip
\item{Fig.~3:}Energy per baryon as a function of $A$ for various choices of
$\epsilon_b$ and $m_s$, including some for which bulk strange matter is
unstable.
\medskip
\item{Fig.~4:}The ratio of radial strangeness density to radial total matter
density for $A=14,\,36,\,100$, showing
the depletion of strangeness near the boundary of the bag.  Here $\epsilon_b =
950$~MeV, $m_s=150$~MeV.
\par
\vfill
\end